\documentclass[12pt, draftclsnofoot, onecolumn]{IEEEtran}
\usepackage{amsmath,amsfonts}
\usepackage{amssymb}
\usepackage{array}
\usepackage[caption=false,font=normalsize,labelfont=sf,textfont=sf]{subfig}
\usepackage{textcomp}
\usepackage[nolist]{acronym}
\begin{acronym}
  \acro{iot}[IoT]{internet of things}
  \acro{bep}[BEP]{bit error probability}
  \acro{ber}[BER]{bit error rate}
  \acro{awgn}[AWGN]{additive white Gaussian noise}
  \acro{snr}[SNR]{signal-to-noise ratio}
  \acro{nd-noma}[ND-NOMA]{noise-domain non-orthogonal multiple access}
  \acro{oodn}[OODN]{on-off digital noise}
  \acro{dbn}[DBN]{differential binary noise}
  \acro{ook}[OOK]{on-off keying}
  \acro{csi}[CSI]{channel state information}
  \acro{tnm}[TNM]{thermal noise modulation}
  \acro{bpsk}[BPSK]{binary phase shift keying}
\acro{bfsk}[BFSK]{binary frequency shift keying}
  \acro{los}[LoS]{line-of-sight}
  \acro{nlos}[NLoS]{non-line-of-sight}
  \acro{noisemod}[NoiseMod]{noise modulation}
  \acro{rf}[RF]{radio-frequency}
  \acro{pdf}[PDF]{probability density function}
  \acro{mmwave}[mmWave]{millimeter wave}  
  \acro{lut}[LUT]{lookup-table}
  \acro{adc}[ADC]{analog-to-digital converter}
  \acro{dac}[DAC]{digital-to-analog converter}
  \acro{clt}[CLT]{central limit theorem}
  \acro{llr}[LRT]{likelihood-ratio test}
  \acro{mse}[MSE]{mean‐square error}
  \acro{pll}[PLL]{phase-locked loop}
  \acro{dsp}[DSP]{digital signal processor}
  \acro{simo}[SIMO]{single-input multiple-output}
  \acro{siso}[SISO]{single-input single-output}
  \acro{miso}[MISO]{multiple-input single-output}
  \acro{egc}[EGC]{equal-gain combining}
  \acro{mrc}[MRC]{maximal-ratio combining}
  \acro{iid}[i.i.d.]{independent and identically distributed}
\end{acronym}
\usepackage{stfloats}
\usepackage{soul}
\usepackage{url}
\usepackage{verbatim}
\usepackage{graphicx}
\usepackage[utf8]{inputenc}
\usepackage{booktabs}
\usepackage{balance}
\usepackage{xcolor}
\usepackage{microtype}
\usepackage{bm}
\usepackage{algorithm}
\usepackage[noend]{algpseudocode}
\usepackage{pifont}
\usepackage[table]{xcolor}
\newcommand{\Real}{\mathfrak{R}}
\newcommand{\Eu}{\mathcal{E}_u}

\usepackage{tabularx}
\newcommand{\CN}{\mathcal{CN}}

\newcommand{\rev}[1]{#1}
\newenvironment{revblock}{}{}

\begin{document}

\title{Receive Diversity for Differential Binary Noise Modulation}

\author{Paulo~V.~B.~Tomé, André~A.~dos~Anjos, Hugerles~S.~Silva,
Daniel~C.~Araújo, Robson~D.~Vieira, and Ertugrul~Basar}

\maketitle

\begin{abstract}
Differential binary noise (DBN) modulation encodes information in the polarity transition of a reused noise realization, dispensing channel state information and carrier-phase recovery. This letter proposes single-input multiple-output (SIMO) reception for DBN via post-correlation decision-statistic combining, followed by one zero-threshold decision. \rev{The combined statistic is a Hermitian quadratic form in complex Gaussian vectors, yielding an exact conditional bit error probability~(BEP) with no Gaussian approximation. Averaging it over $\kappa$-$\mu$ fading and the reused energy gives the exact BEP for arbitrary weights. Because that energy is shared, its deep fades are common to every branch, and the observation length caps the diversity order at $\min(M\mu,N)$. The deflection-optimal weights follow in closed form and interpolate soft combining and power weighting.} Measured 65\,GHz indoor \ac{nlos} results quantify the trade-off between spatial resources and observation length.
\end{abstract}


\section{Introduction}
\acresetall

\Ac{iot} and machine-type communication systems must support
many low-power devices under tight energy and complexity constraints~\cite{Wang},
which motivates keeping the transmitter simple and shifting processing to the
gateway, so as to avoid the circuit-power cost of carrier synchronization,
channel estimation, and coherent demodulation~\cite{Kapetanovic_2022,Basar_2023}.
Noise-based modulation suits this regime by embedding information in the
statistical behavior of random waveforms rather than in deterministic
constellation points. Representative schemes include
variance-switching \ac{noisemod}~\cite{Basar_2024},
\ac{oodn}~\cite{OriginalPaper,oodn_exact}, and pilot-aided, multi-user,
spreading-based, and Gaussian-mixture variants~\cite{Shen_2025,Yapici_2024,%
NoiseModSpread,NoiseMod3D}. Within this family,
\ac{dbn}~\cite{dbn_2026} reuses one noise realization across consecutive symbols and
encodes information in their polarity transition, so a receiver decides from the
sign of an inter-symbol correlation statistic, with no \ac{csi}, noise variance estimation, carrier phase recovery, or adaptive
thresholding. Its reliability, however, is tied to the number of samples per bit
$N$, since a larger $N$ buys processing gain but lengthens the observation
interval and raises the memory and correlation cost per bit.

This raises a natural question: can spatial diversity replace part of the
temporal processing gain while preserving the noncoherent receiver? The question
matters for \ac{iot}, where devices carry one transmit antenna whereas gateways carry
$M$ receive antennas. Rather than combining complex received samples as in
\ac{egc} or \ac{mrc}, the receiver
proposed here forms one \ac{dbn} decision statistic per branch and combines these
real-valued statistics through a generalized weighting rule, recovering soft
combining, \ac{siso} operation, and arbitrary weighted
combining as particular cases.

\rev{The answer is bounded in a way that a single-antenna analysis cannot reveal.
Because the transmitter reuses one realization for every branch, the energy of
that realization is shared by all antennas and cannot be averaged out by adding
more of them, so the diversity order is capped by the observation length rather
than by the array size.}

This paper contributes a generalized \ac{simo}-\ac{dbn}
receiver based on weighted noncoherent decision-statistic combining, which
preserves the original transmitter and zero-threshold detector. \rev{Recognizing
the combined statistic as a Hermitian quadratic form in complex Gaussian
vectors~\cite{Turin_1960}
yields an exact conditional \ac{bep}, valid for every $N$
and free of any Gaussian approximation, from which the exact
average \ac{bep} over generalized $\kappa$-$\mu$ fading follows, evaluated under
measured 65\,GHz indoor \ac{mmwave} \ac{nlos} conditions. An asymptotic analysis then establishes the diversity
order $\min(M\mu,N)$ with its exact leading constant, in which $\mu$ is the
number of multipath clusters of the fading model. Furthermore, the
deflection-optimal combining weights follow in closed form, interpolating soft
combining and power weighting according to the branch \ac{snr}.}

\section{SIMO-DBN System Model}
\label{sec:system}

\begin{figure}[!t]
\centering
\includegraphics[width=\textwidth]{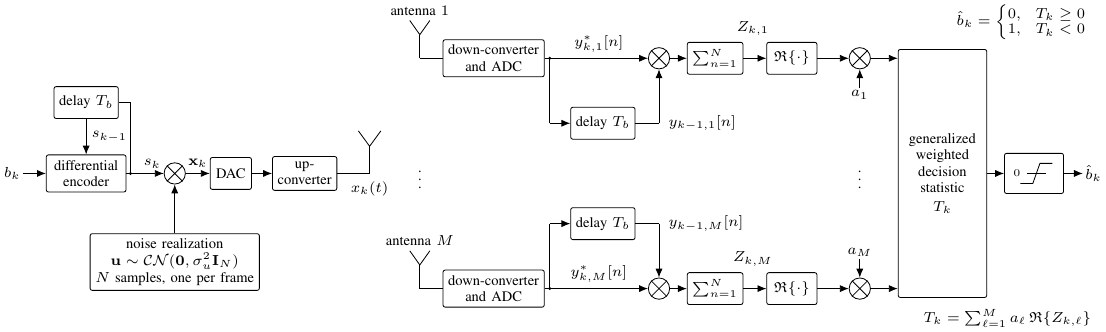}
\caption{Proposed \ac{dbn} transmitter and \ac{simo} receiver architecture. A single noise realization $\mathbf{u}$ is generated once per frame and reused by every receive branch.}
\label{fig:estrutura}
\end{figure}

Fig.~\ref{fig:estrutura} illustrates the proposed \ac{simo}-\ac{dbn} system. The input bits are differentially encoded and modulate a reused complex Gaussian noise realization, which propagates through $M$ independent fading branches, each corrupted by \ac{awgn}. Consider a binary sequence $b_k \in \{0,1\}$ transmitted during the $k$-th bit interval $[(k-1)T_b,kT_b)$, where $T_b$ denotes the bit duration. As in DBN~\cite{dbn_2026}, the bits are differentially encoded into a bipolar sequence $s_k \in \{+1,-1\}$, such that
\begin{equation}
s_k = s_{k-1}(1 - 2b_k), \qquad
d_k \triangleq s_k s_{k-1} = 1 - 2b_k,
\label{eq:diff}
\end{equation}
with an arbitrary initial condition $s_0\in\{+1,-1\}$. Thus, the differential symbol is $d_k=+1$ for $b_k=0$ and $d_k=-1$ for $b_k=1$. Within a frame, a single realization
$\mathbf{u} \sim \CN(\mathbf{0}, \sigma_u^2 \mathbf{I}_N)$ is generated and reused across consecutive intervals, giving the transmitted vector $\mathbf{x}_k = s_k \mathbf{u}$, with $\mathbf{I}_N$ being the $N \times N$ identity matrix and $\sigma_u^2$ denoting the per-sample variance of the noise realization. Its instantaneous energy $\Eu \triangleq \|\mathbf{u}\|^2$ follows a Gamma distribution with shape $N$ and scale $\sigma_u^2$. \rev{Normalizing it as $\varepsilon \triangleq \Eu/\sigma_u^2$ removes that scale from the analysis and gives
\begin{equation}
f_{\varepsilon}(\varepsilon)=
\frac{\varepsilon^{\,N-1}\exp(-\varepsilon)}{\Gamma(N)},
\qquad \varepsilon\ge 0,
\label{eq:Eu_pdf}
\end{equation}
which is the form used throughout.}

With a single transmit antenna and $M$ receive antennas, the $\ell$-th branch observes $\mathbf{y}_{k,\ell}=h_{k,\ell}\mathbf{x}_k+\mathbf{w}_{k,\ell}$, $\ell=1,\ldots,M$, where $h_{k,\ell}$ is the complex fading coefficient and $\mathbf{w}_{k,\ell}\sim\CN(\mathbf{0},\sigma_w^2\mathbf{I}_N)$, with $\sigma_w^2$ being the noise variance per complex sample, is an \ac{awgn} vector independent across antennas, time, and symbols. Adopting the standard two-symbol block-fading assumption of differential detection, $h_{k,\ell}\approx h_{k-1,\ell}\triangleq h_\ell$, valid when the symbol duration is shorter than the channel coherence time, the combining is carried out entirely at the decision-statistic level, keeping channel and combining independent.

\section{Generalized Weighted Decision Statistic}
\label{sec:stat}

At the $\ell$-th receive antenna, the \ac{dbn} differential correlation statistic is
\begin{equation}
Z_{k,\ell} \triangleq \mathbf{y}_{k,\ell}^{\mathsf{H}}\mathbf{y}_{k-1,\ell}
= \sum_{n=1}^{N} y_{k,\ell}^{*}[n]\,y_{k-1,\ell}[n],
\label{eq:Z}
\end{equation}
where $(\cdot)^{\mathsf{H}}$ denotes the Hermitian transpose and $(\cdot)^{*}$ denotes the complex conjugate. The proposed receiver combines these branch statistics according to an arbitrary weighting coefficient $a_\ell$, yielding the decision variable
\begin{equation}
T_k \triangleq \sum_{\ell=1}^{M} a_\ell\,\Real\{Z_{k,\ell}\},
\qquad \mathbf{a} = [a_1,\dots,a_M].
\label{eq:Tk}
\end{equation}
The final decision follows the \ac{dbn} zero-threshold rule, $\hat{b}_k=0$ for $T_k\geq0$ and $\hat{b}_k=1$ otherwise. Unlike \ac{egc} and \ac{mrc}, which combine phase-aligned complex samples, the proposed receiver operates entirely at the decision-statistic level, requiring no carrier-phase recovery, channel-amplitude estimation, or noise-variance estimation, and thus preserving the noncoherent \ac{dbn} structure. Soft combining is the particular case $a_\ell = 1$ for all $\ell$, while other weighting rules leave the analytical framework unchanged.

Substituting $\mathbf{x}_k = s_k\mathbf{u}$ and $\mathbf{y}_{k,\ell} = h_\ell \mathbf{x}_k + \mathbf{w}_{k,\ell}$ into \eqref{eq:Z} gives
\begin{align}
Z_{k,\ell}
&= |h_\ell|^2 d_k \Eu
 + h_\ell^{*} s_k \mathbf{u}^{\mathsf{H}}\mathbf{w}_{k-1,\ell} \notag \\
&\quad + h_\ell s_{k-1}\mathbf{w}_{k,\ell}^{\mathsf{H}}\mathbf{u}
 + \mathbf{w}_{k,\ell}^{\mathsf{H}}\mathbf{w}_{k-1,\ell},
\label{eq:expand}
\end{align}
so that, taking the real part and summing over the $M$ antennas as in \eqref{eq:Tk},
\begin{equation}
T_k = d_k \Eu \sum_{\ell=1}^{M} a_\ell |h_\ell|^2
    + \sum_{\ell=1}^{M} a_\ell \xi_{k,\ell},
\label{eq:central}
\end{equation}
where $\xi_{k,\ell}$ collects the two signal-noise cross terms and the noise-noise term, and the combined instantaneous channel power gain is $G_M \triangleq \sum_{\ell=1}^{M} a_\ell |h_\ell|^2$. The useful term depends on $|h_\ell|^2$ and not on the channel phase, as a direct consequence of the product $h_\ell^{*}h_\ell$ arising from differential correlation, so phase rotations common to two adjacent symbols do not affect the sign of the useful decision component. The weighting vector is treated as deterministic throughout.

\section{Performance Analysis}
\label{sec:bep}

\begin{revblock}

The combined decision statistic $T_k$ in \eqref{eq:Tk} needs no distributional
approximation. Grouping the two
symbols of each sample into $\mathbf{v}_{\ell,n}=[y_{k,\ell}[n],\,
y_{k-1,\ell}[n]]^{\mathsf{T}}$, \eqref{eq:Tk} becomes a Hermitian quadratic form
in complex Gaussian vectors, whose characteristic function
$\phi_T(\omega)\triangleq\mathbb{E}[\exp(\jmath\omega T_k)]$, in which $\omega$
is the transform variable, is
classical. For a complex Gaussian vector of mean $\mathbf{m}$ and covariance
$\bm{\Sigma}$, it reads $\exp[\jmath\omega\mathbf{m}^{\mathsf{H}}(\mathbf{I}-
\jmath\omega\bm{\Sigma}\mathbf{F})^{-1}\mathbf{F}\mathbf{m}]/\det(\mathbf{I}-
\jmath\omega\bm{\Sigma}\mathbf{F})$~\cite{Turin_1960}. Here, $\mathbf{F}$ is a Hermitian matrix with eigenvalues $\pm\frac{1}{2}$ and eigenvectors $[1,\pm1]^{\mathsf{T}}/\sqrt{2}$;
under $d_k=-1$ the mean lies entirely along the $-\frac{1}{2}$ one, so with
soft combining the product over $\ell$ and $n$ collapses to
$\phi_T(\omega)=\exp[-\jmath\omega\Eu G_M/(1+\jmath\omega\sigma_w^2/2)]/
(1+\omega^2\sigma_w^4/4)^{MN}$, which depends on the reused energy and on the
channel only through the single scalar
\begin{equation}
x \triangleq \frac{\Eu\,G_M}{\sigma_w^{2}} .
\label{eq:x}
\end{equation}
With equal weights every branch contributes the same factor, and the scale of
$\omega$ is immaterial because only the sign of $T_k$ decides. Then $\phi_T$ is
that of binary differential phase-shift keying with $L=MN$ diversity branches
and total \ac{snr} $x$. Inverting $\phi_T$ yields the classical
result~\cite[Sec.~13.4]{Proakis_2008},
\begin{equation}
P_e(x)=\frac{e^{-x}}{2^{2L-1}}\sum_{k=0}^{L-1} \beta_k\,x^{k},
\qquad
\beta_k=\frac{1}{k!}\sum_{n=0}^{L-1-k}\binom{2L-1}{n} .
\label{eq:cond_exact}
\end{equation}

Two physical remarks follow. First, $x$ in \eqref{eq:x} is the total energy
collected by the $MN$ correlator taps, normalized by the noise variance, so the conditional error depends on $M$
and $N$ only through their product, each antenna and each sample contributing
one more tap to the same sum.
Second, the differential receiver correlates two noisy copies of the same
waveform rather than a waveform against a clean template, which is why $L=MN$
appears both in the number of branches and in the noise-noise floor of
\eqref{eq:cond_exact} instead of only in the useful term. Unequal weights scale
the factor that each receive branch contributes to $\phi_T$ by its own $a_\ell$,
so the branches no longer share a common scale in $\omega$ and
\eqref{eq:cond_exact} no longer applies. The conditional \ac{bep} then follows
from the same $\phi_T$ by the Gil-Pelaez inversion~\cite{GilPelaez_1951}, which
recovers $\Pr\{T_k<0\}$ from a characteristic function through a single real
integral. Unequal weights leave the diversity order untouched, since the density of
$\sum_\ell a_\ell|h_\ell|^2$ still vanishes as $g^{M\mu-1}$ for any strictly
positive weights and only its constant changes, so weighting buys array gain and
not diversity. Unless otherwise stated, $a_\ell=1$ in what follows, and
Section~\ref{sec:bep}-C returns to the general case.
\end{revblock}

\subsection{Average BEP}

Let $\bar\gamma \triangleq \sigma_u^2/\sigma_w^2$ denote the average per-sample
\ac{snr}, \rev{so that $x=\bar\gamma\,\varepsilon\,G_M$, with $\varepsilon$
distributed as in \eqref{eq:Eu_pdf}}. Since the transmitted energy and the fading process are
independent, the average \ac{bep} is the two-dimensional integral
\begin{equation}
\bar P_e = \int_0^{\infty}\!\!\int_0^{\infty}
P_e\!\left(\bar\gamma\,\varepsilon\,g\right) f_{\varepsilon}(\varepsilon)\,
f_{G_M}(g)\,\mathrm{d}\varepsilon\,\mathrm{d}g .
\label{eq:avg_general}
\end{equation}

\rev{The inner average is available in closed form. Since \eqref{eq:cond_exact}
is a polynomial in $x$ times $e^{-x}$, and $x=\bar\gamma\varepsilon g$ is linear
in $\varepsilon$, each term meets the Gamma density of \eqref{eq:Eu_pdf} in a
single integral,
\begin{equation*}
\int_0^{\infty}\!\varepsilon^{N-1+k}e^{-(1+c)\varepsilon}\,\mathrm{d}\varepsilon
=\frac{\Gamma(N+k)}{(1+c)^{N+k}},
\qquad c\triangleq\bar\gamma\,g,
\end{equation*}
so that averaging \eqref{eq:cond_exact} over the reused energy gives
\begin{equation}
P_e(g)=\frac{1}{2^{2L-1}}\sum_{k=0}^{L-1}\beta_k\,
\frac{\Gamma(N+k)}{\Gamma(N)}\,\frac{c^{k}}{(1+c)^{N+k}} .
\label{eq:cond_g}
\end{equation}
Expression~\eqref{eq:cond_g} is exact rather than a quadrature, and it reduces
\eqref{eq:avg_general} to the single integral
$\bar P_e=\int_0^{\infty}P_e(g)f_{G_M}(g)\,\mathrm{d}g$. It returns $1/2$ as
$c\to0$, since only the $k=0$ term survives and
$\beta_0=2^{2L-2}$. As in \eqref{eq:x}, the branch gain enters only through $c$,
the average post-correlation \ac{snr} per unit of reused energy.}

For \ac{iid} $\kappa$-$\mu$ branches
with common average power $\Omega=\mathbb{E}[|h_\ell|^2]$, in which
$\mathbb{E}[\cdot]$ is the expectation operator, and for soft combining, $G_M$ is the power of an
equivalent $\kappa$-$\mu$ random variable with $\kappa_{\text{eq}}=\kappa$,
$\mu_{\text{eq}}=M\mu$, and $\Omega_{\text{eq}}=M\Omega$~\cite{Yacoub_2007_kappa},
whose power-domain \ac{pdf} is
\begin{align}
f_{G_M}(g) &= \frac{\mu_{\text{eq}}(1+\kappa_{\text{eq}})^{\frac{\mu_{\text{eq}}+1}{2}}
g^{\frac{\mu_{\text{eq}}-1}{2}}}
{\kappa_{\text{eq}}^{\frac{\mu_{\text{eq}}-1}{2}}
 e^{\mu_{\text{eq}}\kappa_{\text{eq}}}\,\Omega_{\text{eq}}^{\frac{\mu_{\text{eq}}+1}{2}}}
\exp\!\left(-\frac{\mu_{\text{eq}}(1+\kappa_{\text{eq}})g}{\Omega_{\text{eq}}}\right)
\notag\\
&\quad\times I_{\mu_{\text{eq}}-1}\!\left(2\mu_{\text{eq}}
\sqrt{\frac{\kappa_{\text{eq}}(1+\kappa_{\text{eq}})g}{\Omega_{\text{eq}}}}\right),
\quad g\geq 0,
\label{eq:kmu_pdf}
\end{align}
in which $I_\nu(\cdot)$ is the modified Bessel function of the first kind and
order $\nu$, $\kappa$ is the power ratio between the dominant component and the
scattered waves, and $\mu$ is associated with the number of multipath clusters.
Setting $\kappa=0$ and $\mu=1$ recovers Rayleigh fading, $\mu=1$ with $\kappa>0$
gives Rician fading, and $\kappa\to0$ gives Nakagami-$m$ with $m=\mu$. 

\rev{Applying $z=\mu_{\text{eq}}(1+\kappa_{\text{eq}})g/\Omega_{\text{eq}}$ to the remaining integral
turns the $\kappa$-$\mu$ density into a Laguerre weight and gives
\begin{align}
\bar P_e &\approx
\frac{(\kappa_{\text{eq}}\mu_{\text{eq}})^{\frac{1-\mu_{\text{eq}}}{2}}}{e^{\mu_{\text{eq}}\kappa_{\text{eq}}}}
\sum_{i=1}^{n_1} w_i\,
z_i^{\frac{\mu_{\text{eq}}-1}{2}}
I_{\mu_{\text{eq}}-1}\!\left(2\sqrt{\kappa_{\text{eq}}\mu_{\text{eq}}z_i}\right)
\notag\\
&\quad\times
P_e\!\left(\frac{\bar\gamma\,z_i\,\Omega_{\text{eq}}}{\mu_{\text{eq}}(1+\kappa_{\text{eq}})}\right),
\label{eq:kmu_bep}
\end{align}
with $z_i$ and $w_i$ being the Laguerre roots and weights, respectively, and
$n_1$ being the number of nodes.
Expression~\eqref{eq:kmu_bep} is accurate while the average is
dominated by channel gains of the order of $\mathbb{E}[G_M]$, which is where the
Laguerre nodes lie.
}

\begin{revblock}
\subsection{Diversity Order}

Let $Y\triangleq \varepsilon\,G_M$, so $x=\bar\gamma Y$. Near the origin the
densities \eqref{eq:Eu_pdf} and \eqref{eq:kmu_pdf} behave as $\varepsilon^{N-1}$
and $g^{\mu_{\text{eq}}-1}$, so $f_Y(y)$ behaves as $C\,y^{d-1}$, with $d$ the
smaller of the two exponents and $C$ a constant fixed below, giving
\begin{equation}
d=\min\left(\mu_{\text{eq}},\,N\right)=\min\left(M\mu,\,N\right)
\label{eq:div}
\end{equation}
and, after rescaling $u=\bar\gamma y$ in \eqref{eq:avg_general},
\begin{equation}
\bar P_e \;\to\; C\,K_d\,\bar\gamma^{-d},
\qquad
K_d=\int_0^{\infty} u^{d-1}P_e(u)\,\mathrm{d}u .
\label{eq:asym}
\end{equation}
The coefficient $C$ follows the same comparison. For $\mu_{\text{eq}}<N$ the
fading sets the exponent and $C=B\,\Gamma(N-\mu_{\text{eq}})/\Gamma(N)$, with
$B$ being the small-argument coefficient of \eqref{eq:kmu_pdf}. For
$\mu_{\text{eq}}>N$ the reused energy sets it and
$C=\mathbb{E}[G_M^{-N}]/\Gamma(N)$, with $\mathbb{E}[G_M^{-N}]$ the $N$-th
negative moment of the combined gain under \eqref{eq:kmu_pdf}. Each form holds
where its own factor converges, $\Gamma(N-\mu_{\text{eq}})$ below the crossing
and $\mathbb{E}[G_M^{-N}]$ above it.

At high \ac{snr} an error requires the useful term of \eqref{eq:central} to
collapse, and only two things can make it collapse. The first is that all $M$
branches fade together. They fade independently, so $\Pr\{G_M<t\}$ falls as
$t^{\mu_{\text{eq}}}$ for small $t$ and every antenna added makes it rarer. The
second is that the reused realization comes out weak, and here antennas do not
help. The same $\mathbf{u}$ multiplies every branch, so one small $\Eu$ wipes
out all $M$ of them at once and $\Pr\{\varepsilon<t\}$ falls as $t^{N}$ no
matter how large the array. Either event alone produces the error, so the more
likely of the two sets the slope, which is \eqref{eq:div}. Past $M\mu=N$ the
weak realization is the more likely one, and extra antennas only raise the
received power.

\subsection{Combining Weight Design}

Write $\gamma_\ell \triangleq \Eu|h_\ell|^2/\sigma_w^2$ for the per-branch
post-correlation \ac{snr}. Conditioned on $\Eu$ and on the branch gains,
\eqref{eq:central} has mean $d_k\sum_\ell a_\ell\gamma_\ell$ and variance
$\sum_\ell a_\ell^2(\gamma_\ell+N/2)$ in units of $\sigma_w^4$, in which
$\gamma_\ell$ comes from the two cross terms of \eqref{eq:expand} and $N/2$ from
its noise-noise term $\mathbf{w}_{k,\ell}^{\mathsf{H}}\mathbf{w}_{k-1,\ell}$.
Maximizing the deflection
$\Psi(\mathbf{a})=\mathbf{a}^{\mathsf{T}}\mathbf{p}/(\mathbf{a}^{\mathsf{T}}
\mathbf{D}\mathbf{a})^{1/2}$, with $p_\ell=\gamma_\ell$ and
$\mathbf{D}=\mathrm{diag}(\gamma_\ell+N/2)$, is a generalized Rayleigh
quotient, solved by $\mathbf{a}^\star\propto\mathbf{D}^{-1}\mathbf{p}$ with
maximum $\Psi^\star=(\mathbf{p}^{\mathsf{T}}\mathbf{D}^{-1}
\mathbf{p})^{1/2}$, that is
\begin{equation}
a_\ell^{\star}\propto\frac{\gamma_\ell}{\gamma_\ell+N/2},
\qquad
\Psi^{\star}=\left[\sum_{\ell=1}^{M}\frac{\gamma_\ell^{2}}{\gamma_\ell+N/2}\right]^{1/2}.
\label{eq:weights}
\end{equation}
The denominator of \eqref{eq:weights} is the variance contributed by a branch, a
signal-noise term proportional to $\gamma_\ell$ plus a noise-noise term
proportional to $N/2$, so a branch is down-weighted only while its own
noise-noise floor dominates its useful energy. The rule gives $a_\ell\to1$ for
$\gamma_\ell\gg N/2$, recovering soft combining, and $a_\ell\propto|h_\ell|^2$
for $\gamma_\ell\ll N/2$, the square-law rule.

Two properties make \eqref{eq:weights} practical. First, $\mathbf{D}$ is
diagonal, so the weights decouple across branches and no matrix inversion is
needed. Second, $\mathbf{u}$ and $h_\ell$ stay constant over a frame, and with
them $\gamma_\ell$, so the weights are computed once and reused for the $K$
decisions of that frame, at a cost negligible against the $MN$ products per bit.

The two implementations differ only in how $\gamma_\ell$ is obtained.
Evaluating \eqref{eq:weights} with the exact $\gamma_\ell$ requires $\Eu$ and
$|h_\ell|^2$ separately, which no noncoherent receiver has, so that version is a
benchmark rather than a receiver. A receiver instead needs $\gamma_\ell$ only in
absolute terms, and the received energy already carries it, since
\begin{equation}
\mathbb{E}\!\left[\|\mathbf{y}_{k,\ell}\|^2\right]
= \Eu|h_\ell|^2 + N\sigma_w^2
= \sigma_w^2\left(\gamma_\ell + N\right).
\label{eq:energy}
\end{equation}
Inverting \eqref{eq:energy} with the frame-averaged received energy
$\hat{E}_\ell$ in place of the expectation of $\|\mathbf{y}_{k,\ell}\|^2$ gives
\begin{equation}
\hat\gamma_\ell=\max\!\left(\frac{\hat{E}_\ell}{\sigma_w^2}-N,\;0\right),
\qquad
\hat{E}_\ell=\frac{1}{K+1}\sum_{k=0}^{K}\|\mathbf{y}_{k,\ell}\|^2 ,
\label{eq:gammahat}
\end{equation}
which fed back into \eqref{eq:weights} yields a blind rule needing no \ac{csi}, no
pilots and no phase reference, at the cost of knowing the receiver noise
floor. Both versions then
combine exactly as in \eqref{eq:Tk} and decide on the same zero threshold, and
Algorithm~\ref{alg:rx} collects the blind one end to end.
\end{revblock}

\begin{algorithm}[!t]
\caption{\Ac{simo}-\ac{dbn} reception of one frame}
\label{alg:rx}
\small
\begin{algorithmic}[1]
\Require $\{\mathbf{y}_{k,\ell}\}$, $k=0,\dots,K$, $\ell=1,\dots,M$;
         noise floor $\sigma_w^2$, window $N$
\Ensure decoded bits $\hat{b}_1,\dots,\hat{b}_K$
\For{$\ell=1$ to $M$}
  \State $\hat{E}_\ell \gets \frac{1}{K+1}\sum_{k=0}^{K}\|\mathbf{y}_{k,\ell}\|^{2}$,\quad
         $\hat\gamma_\ell \gets \max(\hat{E}_\ell/\sigma_w^2-N,\,0)$
         \Comment{\eqref{eq:gammahat}}
  \State $a_\ell \gets \hat\gamma_\ell/(\hat\gamma_\ell+N/2)$
         \Comment{\eqref{eq:weights}, $a_\ell\gets1$ is soft combining}
\EndFor
\For{$k=1$ to $K$}
  \For{$\ell=1$ to $M$}
    \State $Z_{k,\ell} \gets \mathbf{y}_{k,\ell}^{\mathsf{H}}\mathbf{y}_{k-1,\ell}$
           \Comment{\eqref{eq:Z}, no phase reference}
  \EndFor
  \State $T_k \gets \sum_{\ell=1}^{M} a_\ell\,\Real\{Z_{k,\ell}\}$
         \Comment{\eqref{eq:Tk}}
  \State $\hat{b}_k \gets 0$ \textbf{if} $T_k\geq0$ \textbf{else} $1$
         \Comment{fixed zero threshold, $\hat{b}_k$ from \eqref{eq:diff}}
\EndFor
\end{algorithmic}
\end{algorithm}

\section{Numerical Results and Discussion}
\label{sec:numerical}

\begin{revblock}
\begin{figure}[!t]
    \centering
    \includegraphics[width=0.95\linewidth]{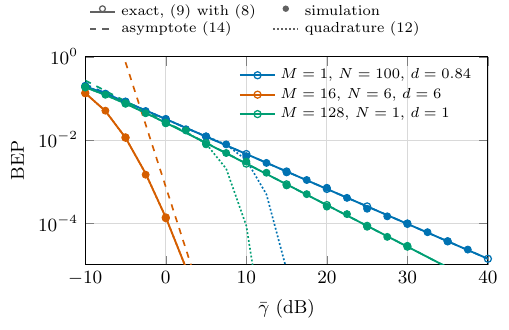}
    \caption{\Ac{bep} versus $\bar\gamma$ under measured 65\,GHz indoor \ac{nlos} fading, $\kappa=1.08$ and $\mu=0.84$, at a fixed budget $MN\approx100$.}
    \label{fig:nlos}
\end{figure}

\begin{figure}[!t]
    \centering
    \includegraphics[width=0.95\linewidth]{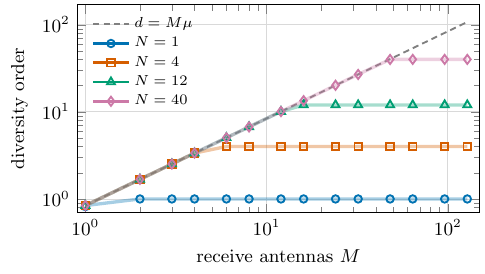}
    \caption{High-\ac{snr} slope of \eqref{eq:avg_general} versus $M$ under measured 65\,GHz indoor \ac{nlos} fading, $\kappa=1.08$ and $\mu=0.84$. The slope tracks $M\mu$ and then saturates at $N$, confirming \eqref{eq:div}.}
    \label{fig:sweep}
\end{figure}

\begin{table}[!t]
\centering
\caption{Simulation parameters}
\label{tab:params}
\footnotesize
\begin{tabular}{@{}ll@{}}
\toprule
Parameter & Value \\
\midrule
Signaling & \ac{dbn}, differential binary, zero threshold \\
Combining & soft, $a_\ell=1$, except in Fig.~\ref{fig:weights} \\
$M$, $N$ & $1$ to $128$ and $1$ to $100$, $MN\approx100$ \\
Frame length & $100$ symbols per noise realization \\
Fading & $\kappa$-$\mu$, block fading over two symbols \\
65\,GHz indoor \ac{nlos} & $\kappa=1.08$, $\mu=0.84$~\cite{dbn_2026,reis_2019} \\
Branch power, noise variance & $\Omega_\ell=\sigma_w^2=1$, so $\bar\gamma=\sigma_u^2/\sigma_w^2$ \\
\Ac{snr} grid & $-10$ to $40$\,dB, step $2.5$\,dB \\
Stopping rule & $200$ errors, floor $10^{-5}$ \\
Integration of \eqref{eq:avg_general} & log-domain grid, step $0.02$ in $\ln$ \\
Quadrature of \eqref{eq:kmu_bep} & $n_1=64$ generalized Laguerre nodes \\
\bottomrule
\end{tabular}
\end{table}

Table~\ref{tab:params} lists the simulation parameters. The Monte Carlo runs the
full link with no approximation of the decision statistic. Each point
accumulates at least $200$ bit errors, which keeps its relative uncertainty
below $8\%$, and points below $10^{-5}$ are not simulated. Throughout, curves
come from the analysis and markers from simulation.

Fig.~\ref{fig:nlos} shows the \ac{bep} of three configurations under soft combining
at an approximately constant budget $MN\approx100$. Four curve types appear.
Solid lines with open markers are \eqref{eq:avg_general} evaluated with the
exact conditional \ac{bep} \eqref{eq:cond_exact}, and filled markers are the
simulated link. Dashed lines are the asymptote \eqref{eq:asym}, and dotted lines
are the quadrature \eqref{eq:kmu_bep}.

Three observations follow from Fig.~\ref{fig:nlos}. First, the analysis matches
the simulation over the whole range and for every configuration, which validates
\eqref{eq:cond_exact} and \eqref{eq:avg_general} together. Second, the
configurations differ in slope and not only in offset, and those slopes are the
ones \eqref{eq:div} predicts, six for $M=16$ with $N=6$ and one for $M=128$ with
$N=1$. The latter is barely above the single-antenna link, so $127$ extra
antennas move the curve sideways rather than tilting it. Third, the asymptote
\eqref{eq:asym} settles onto the analysis in both slope and constant, to within
$1\%$ beyond $24$\,dB, so it can size a link budget on its own, whereas the
quadrature \eqref{eq:kmu_bep} follows the analysis only up to about $10$\,dB,
for the reason given in Section~\ref{sec:bep}-A.

Fig.~\ref{fig:sweep} plots the high-\ac{snr} slope of \eqref{eq:avg_general}
against $M$ for four observation lengths. Each curve climbs along $M\mu$ and
then flattens, and it flattens at the very $N$ that produced it, so the four
plateaus are the four observation lengths. That is \eqref{eq:div} read off the
figure. Antennas set the diversity order only while $M\mu<N$, and past the
crossing the reused realization sets it, so the ceiling does not move over more
than two decades in $M$. No array size buys diversity beyond that point, which
is what separates \ac{simo}-\ac{dbn} from a receiver whose branches carry
independent waveforms.

\begin{figure}[!t]
    \centering
    \includegraphics[width=0.95\linewidth]{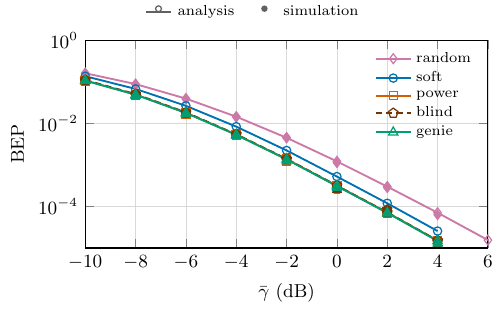}
    \caption{Weighting strategies under measured 65\,GHz indoor \ac{nlos} fading, $M=4$ and $N=25$.}
    \label{fig:weights}
\end{figure}

Fig.~\ref{fig:weights} compares weighting rules for $M=4$ and $N=25$, all
sharing the same statistic and threshold and differing only in $a_\ell$. Soft
combining uses $a_\ell=1$, the square-law (power) rule $a_\ell\propto|h_\ell|^2$, and
the random rule draws $a_\ell$ from the uniform distribution on $[0.05,1]$ once per frame with no
knowledge of the channel. The genie rule evaluates the deflection-optimal
\eqref{eq:weights} with the exact $\gamma_\ell$, making it a benchmark rather
than a receiver, and the blind rule feeds that same expression with the estimate
\eqref{eq:gammahat}, as in Algorithm~\ref{alg:rx}.

Weighting without knowing the branches is worse than not weighting
at all. The random choice falls $1.34$\,dB below soft combining. The genie rule
gains $0.70$\,dB, and the square law is indistinguishable from it, since
the average is carried by deep fades, where $\gamma_\ell\ll N/2$ and
\eqref{eq:weights} degenerates into $a_\ell\propto|h_\ell|^2$. The gain thus
comes from the noise floor rather than from the fading. The blind rule
recovers nearly all of it without \ac{csi}, so \eqref{eq:gammahat} resolves
$\gamma_\ell$ finely. The gains are largely unchanged when the average branch
powers differ, which suggests that weighting draws mainly on the instantaneous
imbalance within a frame rather than on any persistent asymmetry between
antennas.


\end{revblock}

\section{Conclusion}
\label{sec:conclusion}

\begin{revblock}
This paper proposes a receive-diversity scheme for \ac{dbn} modulation based
on generalized noncoherent decision-statistic combining. Recognizing the
combined statistic as a Hermitian quadratic form in complex Gaussian vectors
yields an exact conditional \ac{bep} for every observation length. Averaging it
over $\kappa$-$\mu$ fading and over the reused-energy law gives the exact
\ac{bep}, validated against Monte Carlo of the full link. The asymptotic
analysis establishes the diversity order $\min(M\mu,N)$. The deflection-optimal
combining weights follow in closed form, interpolating soft combining and power
weighting, and a blind version of the same rule attains them without \ac{csi}. \Ac{miso}
and space-time extensions remain as future work.
\end{revblock}

\bibliographystyle{ieeetran}
\bibliography{library}

\end{document}